\documentclass[copyright,creativecommons]{eptcs}
\providecommand{\event}{HPC'10}  
\providecommand{\firstpage}{1}
\usepackage{amssymb}
\usepackage{latexsym}
\usepackage{amsmath}
\usepackage{epsfig}

\newtheorem{Th}{Theorem}[section]

\newtheorem{proposition}[Th]{Proposition}

\newtheorem{definition}[Th]{Definition}

\newcommand{\bit}{\begin{itemize}}
\newcommand{\eit}{\end{itemize}\par\noindent}
\newcommand{\ben}{\begin{enumerate}}
\newcommand{\een}{\end{enumerate}\par\noindent}
\newcommand{\beq}{\begin{equation}}
\newcommand{\eeq}{\end{equation}\par\noindent}
\newcommand{\beqa}{\begin{eqnarray*}}
\newcommand{\eeqa}{\end{eqnarray*}\par\noindent}
\newcommand{\beqn}{\begin{eqnarray}}
\newcommand{\eeqn}{\end{eqnarray}\par\noindent}

\def\bR{\begin{color}{red}}
\def\bB{\begin{color}{blue}}
\def\bM{\begin{color}{magenta}}
\def\bC{\begin{color}{cyan}}
\def\bW{\begin{color}{white}}
\def\bBl{\begin{color}{black}}
\def\bG{\begin{color}{green}}
\def\bY{\begin{color}{yellow}}
\def\e{\end{color}}

\usepackage{ifthen}
\usepackage{tikzfig}
\usetikzlibrary{shapes}
\usetikzlibrary[shapes.geometric]
\pgfdeclarelayer{edgelayer}
\pgfdeclarelayer{nodelayer}
\pgfsetlayers{edgelayer,nodelayer,main}

\tikzstyle{white dot}=[dot,fill=white]
\tikzstyle{none}=[inner sep=0mm]

\tikzstyle{dot}=[inner sep=1mm,fill=black,draw=black,shape=circle]
\tikzstyle{dotpic}=[baseline=-0.25em,shorten <=-0.1mm,shorten >=-0.1mm,scale=0.6]
\tikzstyle{small}=[inner sep=0.4mm]
\tikzstyle{dotpic inline}=[baseline=(current bounding box).south]
\tikzstyle{cdiag}=[baseline=(current bounding box).east,xscale=1.5,-latex]
\tikzstyle{every loop}=[]
\tikzstyle{rn}=[dot]
\tikzstyle{gn}=[dot]
\tikzstyle{bn}=[inner sep=0pt]
\tikzstyle{uploop}=[in=45,out=135,loop]
\tikzstyle{downloop}=[in=-45,out=-135,loop]
\tikzstyle{small dot}=[dot,inner sep=0.4mm]
\tikzstyle{small white dot}=[small dot,fill=white]
\tikzstyle{small gray dot}=[small dot,fill=gray!50]
\tikzstyle{greenbox}=[rectangle,fill=gray!30,draw=gray!50!black,minimum height=7mm,minimum width=7mm]
\tikzstyle{bluebox}=[rectangle,fill=white,draw=gray,minimum height=7mm,minimum width=7mm]
\tikzstyle{cnot}=[fill=white,shape=circle,inner sep=-1.4pt]
\tikzstyle{pt}=[regular polygon,regular polygon sides=3,draw=black,scale=0.75,inner sep=-0.5pt]
\tikzstyle{copt}=[pt,regular polygon rotate=180]
\tikzstyle{tick}=[sloped,rotate=90,font=\small\bf,xshift=0.07mm]
\tikzstyle{white dot}=[dot,fill=white]
\tikzstyle{gray dot}=[dot,fill=gray!50]
\tikzstyle{red dot}=[dot,fill=red!50]%%%%%%%%%%%%%
\tikzstyle{green dot}=[dot,fill=green!50]%%%%%%%%%%%%%
\tikzstyle{gs dot}=[dot,fill=gray]
\tikzstyle{small dotpic}=[dotpic,scale=0.6]
\tikzstyle{mux}=[rectangle,draw=black,scale=0.5,minimum width=1.8cm,minimum height=1cm]

\tikzstyle{square box}=[rectangle,fill=white,draw=black,minimum height=6mm,minimum width=6mm]
\tikzstyle{square gray box}=[rectangle,fill=gray!30,draw=black,minimum height=6mm,minimum width=6mm]
\tikzstyle{dashed box}=[draw=black,dashed,minimum height=12mm,minimum width=12mm,fill=gray!20]
\tikzstyle{box vertex}=[rectangle,draw=black]

\newcommand{\ket}[1]{|#1\rangle}

\title{Three qubit entanglement within graphical Z/X-calculus}
\author{
Bob Coecke and Bill Edwards\thanks{This work supported by BC's EPSRC Advanced Research Fellowship EP/D072786/1,
BE's EPSRC DTA Studentship,  Office of Naval Research Grant N00014-09-1-0248, an FQXi Large Grant,  and EU FP6 STREP QICS.
We thank Ross Duncan and Aleks Kissinger for discussions related to the content of this paper.}
\institute{Oxford University Computing Laboratory, Quantum Group\\
Wolfson Building, Parks Road, OX1 3QD Oxford}
\email{coecke@comlab.ox.ac.uk/wae28@hotmail.com}
}

\def\titlerunning{Three qubit entanglement within graphical Z/X-calculus}
\def\authorrunning{B. Coecke \& B. Edwards}

\begin{document}
\maketitle

%\begin{titlepage}
%\begin{center}
%\textbf{\Large{Computing Laboratory}}
%\end{center}
%\vspace{5mm}
%\begin{center}
%\Large{THREE QUBIT ENTANGLEMENT\\ IN GRAPHICAL Z/X-CALCULUS}
%\end{center}
%\vspace{5mm}
%\begin{center}
%\Large{Bob Coecke and Bill Edwards}
%\end{center}
%\vspace{2mm}
%\begin{center}
%\Large{CS-RR-09-12}
%\end{center}
%\vspace{20mm}
%\begin{center}
%\epsfig{figure=Oxfordlogo.eps,width=130pt,height=150pt}
%\end{center}
%\vspace{20mm}
%\begin{center}
%\Large{Oxford University Computing Laboratory}\\
%\Large{Wolfson Building, Parks Road, Oxford, OX1 3QD}
%\end{center}
%\end{titlepage}

\begin{abstract}
The compositional techniques of categorical quantum mechanics are applied to
analyse 3-qubit quantum entanglement. In particular the graphical calculus of complementary
observables and corresponding phases due to Duncan and one of the authors is used to construct
representative members of the two genuinely tripartite SLOCC classes of 3-qubit entangled states,
GHZ and W. This nicely illustrates the respectively pairwise and global tripartite entanglement found in the W- and
GHZ-class states.  A new concept of supplementarity allows us to characterise inhabitants of the W class within the
abstract diagrammatic calculus; these method extends to  more general multipartite qubit states.
\end{abstract}

\section{Introduction}

The structure of multipartite entanglement has been a subject of much research in recent years. Much work has been done on trying to classify the entanglement in many body states (for example \cite{DVC3qubit, 4qubit, 2x2xn}), and investigating the properties and uses of particular multipartite entangled states, for example graph states in measurement based quantum computing \cite{HEB04}. Three qubit states with genuine tripartite entanglement fall into two SLOCC\footnote{SLOCC: Stochastic Local Operations and Classical Communication.}-classes: one inhabited by the GHZ-state, which is a graph states, and one inhabited by the W-state \cite{DVC3qubit}.  %Still, much remains mysterious. For example there is currently no overarching classification of multipartite entangled states.
But beyond these states, there is little structural understanding of general  multipartite entangled states; even their classification remains mysterious. Such states also remain virtually untapped as resources for quantum information processing protocols.

Most of the work described above has so far employed rather technical arguments using linear algebra. This paper describes some tentative steps towards a different, \emph{compositional} approach, where we view entangled states as being built up from simpler components. This approach springs from the programme, initiated by Abramsky  and one of the authors \cite{CatSem}, to analyse quantum mechanics in terms of symmetric monoidal categories. We assume that the reader is familiar with the basics of this approach; for an introductory account see \cite{ContPhys}. One feature of this programme is that it allows us to use a very intuitive graphical language to describe quantum states and processes, and this is utilised in much of what follows.

Here we show how to build up examples from the W SLOCC class using simple graphical building blocks; the GHZ-state is itself one of the building blocks. These building blocks are categorical structures called \emph{basis structures} \cite{wosums}; in the categorical quantum mechanics programme these provide an abstract counterpart to orthonormal bases. In particular we will use a pair of \emph{complementary} basis structures - this notion was introduced by Duncan  and one of the authors \cite{RedGreen} to model mutually unbiased bases. Their  \em graphical Z/X-calculus \em employed a vivid graphical convention where two complementary basis structures were depicted using green and red dots respectively.

To understand the difference between the GHZ class states and W class states  within the context of abstract categorical quantum mechanics, we introduce a concept of the \em supplementarity \em of certain elements  relative to complementary basis structures; W states then arise in situations of supplementarity. We sketch how this concept leads to distinct subclasses of more general multipartite qubit states.

One important  advantage of the graphical presentation of entangled states is that it is subject to automated reasoning, by means of the {\tt quantomatic} software  \cite{quanto}, developed by Dixon, Duncan, Kissinger and Merry.
\begin{figure}
\begin{center}
\epsfig{figure=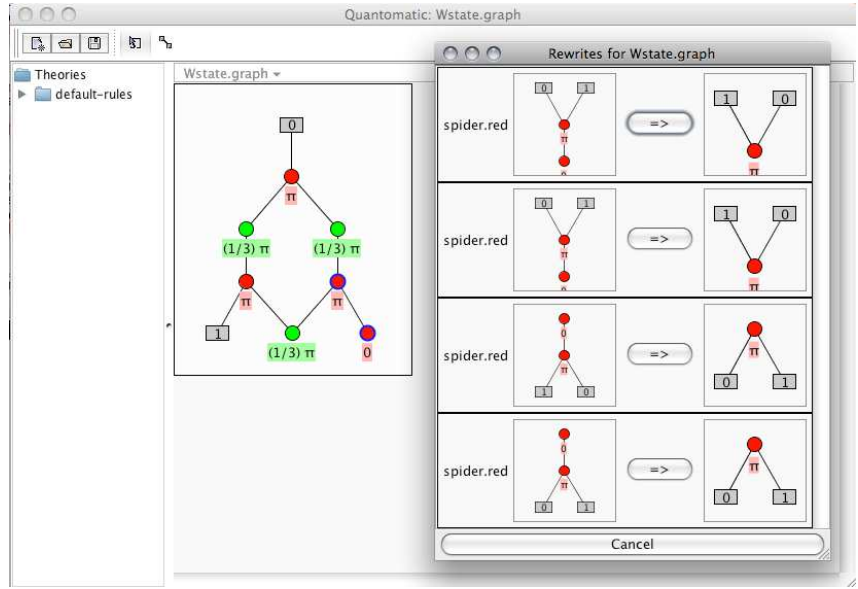,width=320pt}
\caption{Screenshot of  some suggested rewrites in {\tt quantomatic}.}
\end{center}
\end{figure}
Meanwhile, the results in this paper have also led to  a  new graphical calculus, which takes both the GHZ and W state as its primitives  \cite{CK}. Rather than being mediated by the laws of complementarity, this graphical GHZ/W-calculus is mediated by the laws of basic arithmetic  \cite{CKMS}.

This paper is structured as follows: in section \ref{3qubitbackgroundSec} we give a brief introduction to some key aspects of three-qubit entanglement, and section \ref{redgreenbackgroundSec} reviews some essential properties of complementary basis structures. Next, in section \ref{GHZWgraphicallySec} we show (via concrete linear algebra calculations) how certain states from the GHZ and W classes can be built up from the morphisms of a pair of complementary basis structures.
In section \ref{sec:topoclass} we show how one can identify W class states within the graphical calculus by means of the notion of supplementarity.
%Having done this we can use the simple and elegant rules of the graphical language to establish properties of these states which would be rather more tedious to derive using explicit linear algebra.
We conclude with some speculations on whether our methods can be used for the study of general multipartite entanglement.

\section{Background: 3-qubit entanglement}\label{3qubitbackgroundSec}

Entangled states are classified primarily according to the following criterion: If state $\ket{\psi}$ can be transformed into state $\ket{\phi}$ via local operations on the components of the system (we also allow classical communication between the agents acting on the components, so that they can condition their operations on the outcomes of measurements performed by other agents, for example) then $\ket{\psi}$ is more entangled than, or as entangled as $\ket{\phi}$.

In the case of bipartite systems (with components whose state spaces have arbitrary dimension) the entanglement of quantum states has been completely classified, in that there exists a well-defined mathematical criterion for determining whether one state can be transformed into another via LOCC (local operations and classical communication). This leads to the well-known \emph{majorization order} \cite{NieMaj}.

For systems with more than two subsystems (henceforth we will refer to this as \emph{multipartite} entanglement) no such classification has been achieved. However in certain cases a weaker classification has been achieved. This is based on whether one state can be converted into another via LOCC \emph{with some non-zero probability}. In this case we say that the two states are related via SLOCC (\emph{stochastic} local operations and classical communication).

Mathematically SLOCC translates into a very simple condition \cite{DVC3qubit}. For the $n$-partite state $\ket{\psi}$ to be transformable into $\ket{\phi}$ via SLOCC there must exist local linear operations $A_1,A_2,\dots,A_n$ such that:
\begin{equation} \label{sloccconditionEq}
\ket{\phi} = A_1\otimes A_2 \otimes \dots \otimes A_n \ket{\psi}
\end{equation}

%Graphically this condition would be depicted as:
%
%\begin{equation}
%\def\JPicScale{0.8}
%\input{pictures/SLOCC.pst}
%\end{equation}

In practice determining whether this condition is satisfied for two states is not straightforward. However in the case of three qubits the states have been completely classified under SLOCC \cite{DVC3qubit}. There are six classes of states arranged in a hierarchy (in fact a partial order). The states of a given class are all interconvertible under SLOCC. States from higher classes can be converted via SLOCC into states from lower classes.
\[
\begin{tikzpicture}[scale=1.1]
    \begin{pgfonlayer}{nodelayer}
        \node [style=white dot] (0) at (-7, 0.25) {\small GHZ};
        \node [style=white dot] (1) at (-5, 0.25) {\small \ W \  };
        \node [style=white dot] (2) at (-8, -1) {\small A-BC};
        \node [style=white dot] (3) at (-6, -1) {\small B-CA};
        \node [style=white dot] (4) at (-4, -1) {\small C-AB};
        \node [style=white dot] (5) at (-6, -2.25) {\small $\!$A-B-C$\!$};
    \end{pgfonlayer}
    \begin{pgfonlayer}{edgelayer}
        \draw (3) to (5);
        \draw (1) to (4);
        \draw (2) to (5);
        \draw (0) to (3);
        \draw (4) to (5);
        \draw (3) to (1);
        \draw (2) to (1);
        \draw (2) to (0);
        \draw (0) to (4);
    \end{pgfonlayer}
\end{tikzpicture}\]
%\begin{center}
%\def\JPicScale{0.4}
%\input{pictures/tripartitehierarchy.pst}
%\end{center}
The bottom class contains the completely separable states. The middle three classes consist of states where one system is unentangled, while the other two are entangled with each other. There are two classes of true tripartite entanglement, not interconvertible, each named after a particular member state:
\begin{eqnarray}
|\Psi_{GHZ}\rangle & = & \frac{1}{\sqrt{2}} ( |000\rangle + |111\rangle )  \\
|\Psi_{W}\rangle      & = & \frac{1}{\sqrt{3}} ( |011\rangle + |101\rangle + |110\rangle )
\end{eqnarray}

Is there any intuitive difference between the GHZ-class and W-class states? One difference is mostly clearly seen using a measure of entanglement called the \emph{tangle}, introduced in \cite{CKWtangle}. Label the three qubits $A$, $B$ and $C$. Strictly the tangle is defined on mixed states of two qubits. However it is possible to sensibly extend this definition and calculate the tangle, $\tau_{A(BC)}$ for the entanglement between qubit $A$ and qubits $B$ and $C$ viewed as a single 4-dimensional system. The tangle for the entanglement between $A$ and $B$, $\tau_{AB}$ and between $A$ and $C$, $\tau_{AC}$, can also be calculated. The following inequality then always holds:
\begin{equation}
\tau_{A(BC)} \geq \tau_{AB} + \tau_{AC}
\end{equation}
and likewise for permutations of $A$, $B$ and $C$.  Interestingly the quantities:
\begin{eqnarray}
\tau_{A(BC)} - \tau_{AB} - \tau_{AC}\\
\tau_{B(AC)} - \tau_{AB} - \tau_{BC}\\
\tau_{C(AB)} - \tau_{BC} - \tau_{AC}
\end{eqnarray}
are all equal. This quantity is termed the \emph{3-tangle} and is denoted $\tau_{ABC}$. We can now write, for example:
\begin{equation}
\tau_{A(BC)} = \tau_{AB} + \tau_{AC} + \tau_{ABC}
\end{equation}

This seems to say that the entanglement between $A$ and the combined system of $B$ and $C$ consists of the pairwise entanglement of $A$ with $B$ and $A$ with $C$, plus some kind of global tripartite entanglement, quantified by the 3-tangle.

It can be shown \cite{DVC3qubit} that all GHZ-class states have non-zero 3-tangle, while all W states have zero 3-tangle. Thus it would seem that in some sense, the entanglement in W-class states is all pairwise, or local, while  in the case of GHZ-class states, at least some of the entanglement is genuinely global, shared between all three qubits.

\section{Background: Red-Green calculus}\label{redgreenbackgroundSec}

We consider as given a dagger symmetric monoidal category ($\dag$-SMC) \cite{CatSem}, that is, a symmetric monoidal category with an identity-on-objects involutive contravariant functor $f\mapsto f^\dagger$.  We will work  within its corresponding graphical calculus \cite{JoyalStreet}.

In a $\dag$-SMC a \emph{basis structure} is a dagger special commutative Frobenius algebra \cite{wosums} - i.e.~a refinement   of Carboni and Walters' frobenius algebras \cite{CarboniWalters}. It consists of  an internal commutative monoid
\[
(A, \delta:A\rightarrow A\otimes A, \epsilon: A\rightarrow I)
\]
 for which  $\delta$ and $\delta^\dagger$ satisfy the specialness and Frobenius equations, that is:
\[
\begin{tikzpicture}
    \begin{pgfonlayer}{nodelayer}
        \node [style=none] (0) at (-3.5, -1.5) {};
        \node [style=none] (1) at (-1.75, -1.5) {};
        \node [style=none] (2) at (0.5, -1.5) {};
        \node [style=none] (3) at (1.5, -1.5) {};
        \node [style=none] (4) at (3, -1.5) {};
        \node [style=none] (5) at (4, -1.5) {};
        \node [style=green dot] (6) at (-3.5, -1.75) {};
        \node [style=green dot] (7) at (1.5, -2) {};
        \node [style=green dot] (8) at (3.5, -2) {};
        \node [style=none] (9) at (-2.5, -2.25) {=};
        \node [style=none] (10) at (2.5, -2.25) {=};
        \node [style=green dot] (11) at (1, -2.5) {};
        \node [style=green dot] (12) at (3.5, -2.5) {};
        \node [style=green dot] (13) at (-3.5, -2.75) {};
        \node [style=none] (14) at (-3.5, -3) {};
        \node [style=none] (15) at (-1.75, -3) {};
        \node [style=none] (16) at (1, -3) {};
        \node [style=none] (17) at (2, -3) {};
        \node [style=none] (18) at (3, -3) {};
        \node [style=none] (19) at (4, -3) {};
    \end{pgfonlayer}
    \begin{pgfonlayer}{edgelayer}
        \draw (16.center) to (11);
        \draw (7) to (3.center);
        \draw (0.center) to (6);
        \draw (13) to (14.center);
        \draw[bend right=315, looseness=1.50] (6) to (13);
        \draw[bend right=45, looseness=1.50] (6) to (13);
        \draw (8) to (5.center);
        \draw (1.center) to (15.center);
        \draw (8) to (12);
        \draw (8) to (4.center);
        \draw[bend left=15] (11) to (2.center);
        \draw (18.center) to (12);
        \draw (7) to (11);
        \draw[bend left=15, looseness=1.25] (7) to (17.center);
        \draw (12) to (19.center);
    \end{pgfonlayer}
\end{tikzpicture}
\]
where we represented $\delta$, $\delta^\dagger$,  $\epsilon$ and $\epsilon^\dagger$ graphically as:
\[
\begin{tikzpicture}[scale=0.8]
    \begin{pgfonlayer}{nodelayer}
        \node [style=none] (0) at (3, -1.5) {};
        \node [style=none] (1) at (4, -1.5) {};
        \node [style=none] (2) at (6, -1.5) {};
        \node [style=green dot] (3) at (8, -1.75) {};
        \node [style=none] (4) at (10, -1.75) {};
        \node [style=green dot] (5) at (3.5, -2) {};
        \node [style=green dot] (6) at (6, -2.25) {};
        \node [style=none] (7) at (8, -2.5) {};
        \node [style=green dot] (8) at (10, -2.5) {};
        \node [style=none] (9) at (3.5, -2.75) {};
        \node [style=none] (10) at (5.5, -2.75) {};
        \node [style=none] (11) at (6.5, -2.75) {};
    \end{pgfonlayer}
    \begin{pgfonlayer}{edgelayer}
        \draw (3) to (7.center);
        \draw (6) to (10.center);
        \draw (6) to (2.center);
        \draw (5) to (1.center);
        \draw (6) to (11.center);
        \draw (8) to (4.center);
        \draw (5) to (9.center);
        \draw (5) to (0.center);
    \end{pgfonlayer}
\end{tikzpicture}
\]

%We represent the morphisms $\delta$ and $\epsilon$ graphically as:
%\begin{equation}
%\def\JPicScale{0.6}
%\input{pictures/deltaandepsilon.pst}
%\end{equation}

A key result holding for any basis structure is the \emph{spider theorem}  - a high-level abstract account of which is due to Lack \cite{Lack}. This spider theorem essentially allows us, when working with the graphical language, to fuse together any directly connected dots representing $\delta, \delta^\dag, \epsilon$ and $\epsilon^\dag$ from the same basis structure: the theorem guarantees that all morphisms which \emph{look} the same graphically after such a fusing procedure \emph{are} indeed equal. In fact, this property provides an equivalent definition of basis structures, since all the defining axioms of a basis structure are implied by it \cite{ContPhys, RedGreen}.

Among many other things, the spider theorem implies that we have a compact structure \cite{KellyLaplaza}:
\[
\begin{tikzpicture}[scale=0.8]
    \begin{pgfonlayer}{nodelayer}
        \node [style=green dot] (0) at (4, -1.25) {};
        \node [style=none] (1) at (4, -1.25) {};
        \node [style=none] (2) at (8.25, -1.25) {};
        \node [style=none] (3) at (10.25, -1.25) {};
        \node [style=none] (4) at (-1.75, -1.5) {};
        \node [style=none] (5) at (-0.75, -1.5) {};
        \node [style=green dot] (6) at (6.75, -1.5) {};
        \node [style=none] (7) at (-4.25, -1.75) {};
        \node [style=none] (8) at (-3.25, -1.75) {};
        \node [style=green dot] (9) at (1.5, -1.75) {};
        \node [style=none] (10) at (-2.5, -2) {$=$};
        \node [style=green dot] (11) at (-1.25, -2) {};
        \node [style=none] (12) at (2.75, -2) {$=$};
        \node [style=green dot] (13) at (4, -2) {};
        \node [style=none] (14) at (6.25, -2) {};
        \node [style=none] (15) at (7.25, -2) {};
        \node [style=none] (16) at (7.25, -2) {};
        \node [style=none] (17) at (8.25, -2) {};
        \node [style=none] (18) at (9.25, -2) {$=$};
        \node [style=green dot] (19) at (-3.75, -2.25) {};
        \node [style=none] (20) at (1, -2.25) {};
        \node [style=none] (21) at (2, -2.25) {};
        \node [style=none] (22) at (3.5, -2.5) {};
        \node [style=none] (23) at (4.5, -2.5) {};
        \node [style=green dot] (24) at (7.75, -2.5) {};
        \node [style=none] (25) at (-1.25, -2.75) {};
        \node [style=green dot] (26) at (-1.25, -2.75) {};
        \node [style=none] (27) at (6.25, -2.75) {};
        \node [style=none] (28) at (10.25, -2.75) {};
    \end{pgfonlayer}
    \begin{pgfonlayer}{edgelayer}
        \draw (2.center) to (17.center);
        \draw[bend left=45, looseness=1.25] (24) to (16.center);
        \draw (13) to (1.center);
        \draw[bend left=45, looseness=1.25] (19) to (7.center);
        \draw[bend right=45, looseness=1.25] (9) to (20.center);
        \draw[bend left=45, looseness=1.25] (6) to (15.center);
        \draw (11) to (25.center);
        \draw (3.center) to (28.center);
        \draw (11) to (4.center);
        \draw (27.center) to (14.center);
        \draw (13) to (23.center);
        \draw (13) to (22.center);
        \draw[bend left=45, looseness=1.25] (9) to (21.center);
        \draw (11) to (5.center);
        \draw[bend right=45, looseness=1.25] (19) to (8.center);
        \draw[bend right=45, looseness=1.25] (6) to (14.center);
        \draw[bend right=45, looseness=1.25] (24) to (17.center);
    \end{pgfonlayer}
\end{tikzpicture}
\]
Such a compact structure allows one to exchange the roles of inputs to outputs, and vice versa, so we can essentially ignore those roles. When writing equations it suffices to identify open-ended wires in the LHS with those in the RHS.

Compact structure induces
%is coherently defined for all the objects in a $\dagger$-SMC this gives rise to
a covariant involutive \em conjugation \em functor:
\[
\begin{tikzpicture}[scale=1]
    \begin{pgfonlayer}{nodelayer}
        \node [style=none] (0) at (4.25, -1.25) {};
        \node [style=none] (1) at (8.25, -1.25) {};
        \node [style=green dot] (2) at (6.75, -1.5) {};
        \node [style=white dot] (3) at (4.25, -2) {$f$};
        \node [style=none] (4) at (5.25, -2) {$\mapsto$};
        \node [style=none] (5) at (6.25, -2) {};
        \node [style=white dot] (6) at (7.25, -2) {$f^\dagger$};
        \node [style=none] (7) at (7.25, -2) {};
        \node [style=none] (8) at (8.25, -2) {};
        \node [style=green dot] (9) at (7.75, -2.5) {};
        \node [style=none] (10) at (4.25, -2.75) {};
        \node [style=none] (11) at (6.25, -2.75) {};
    \end{pgfonlayer}
    \begin{pgfonlayer}{edgelayer}
        \draw (1.center) to (8.center);
        \draw[bend left=45, looseness=1.25] (9) to (7.center);
        \draw[bend right=45, looseness=1.25] (9) to (8.center);
        \draw (0.center) to (3.center);
        \draw[bend right=45, looseness=1.25] (2) to (5.center);
        \draw (3.center) to (10.center);
        \draw (11.center) to (5.center);
        \draw[bend left=45, looseness=1.25] (2) to (6.center);
    \end{pgfonlayer}
\end{tikzpicture}
\]

In the category \textbf{FdHilb} of finite dimensional Hilbert spaces, linear maps, tensor products and adjoints, basis structures are in bijective correspondence with orthonormal bases \cite{CocPavVic}. Explicitly, for a Hilbert space $\mathcal{H}$, $\delta$ `copies' basis vectors and $\epsilon$ `uniformly erases' them:
\begin{equation}\label{eq:concreteBS}
\delta:\mathcal{H}\rightarrow\mathcal{H}\otimes\mathcal{H}:: \ket{i}\mapsto\ket{ii}\qquad\textrm{and}\qquad
\epsilon:\mathcal{H}\rightarrow\mathbb{C}:: \ket{i}\mapsto 1\,.
\end{equation}
Basis structures act as the abstract counterparts to orthonormal bases, and in analogy to the concrete case, we define the \em basis elements \em of  general basis structures as the elements $\psi:I\to A$ that are self-conjugate comonoid  homomorphisms  \cite{RedGreen}, that is, explicitly:
 \[
\begin{tikzpicture}[scale=1]
    \begin{pgfonlayer}{nodelayer}
        \node [style=none] (0) at (-7.75, -1.5) {};
        \node [style=none] (1) at (-6.75, -1.5) {};
        \node [style=none] (2) at (-5.25, -1.5) {};
        \node [style=none] (3) at (-4.25, -1.5) {};
        \node [style=green dot] (4) at (-2, -1.5) {};
        \node [style=green dot] (5) at (4, -1.5) {};
        \node [style=white dot] (6) at (6.25, -1.5) {$\psi$};
        \node [style=green dot] (7) at (-7.25, -2) {};
        \node [style=none] (8) at (-6.25, -2) {$\stackrel{C_1}{=}$};
        \node [style=none] (9) at (-1.25, -2) {$\stackrel{C_2}{=}$};
        \node [style=none] (10) at (0.5, -2) {\mbox{\rm ``empty picture"}};
        \node [style=none] (11) at (3.5, -2) {};
        \node [style=white dot] (12) at (3.5, -2) {$\psi^\dagger$};
        \node [style=none] (13) at (4.5, -2) {};
        \node [style=none] (14) at (5.25, -2) {$\stackrel{C_3}{=}$};
        \node [style=white dot] (15) at (-5.25, -2.25) {$\psi$};
        \node [style=white dot] (16) at (-4.25, -2.25) {$\psi$};
        \node [style=none] (17) at (-2, -2.5) {};
        \node [style=white dot] (18) at (-2, -2.5) {$\psi$};
        \node [style=none] (19) at (4.5, -2.5) {};
        \node [style=none] (20) at (6.25, -2.5) {};
        \node [style=white dot] (21) at (-7.25, -2.75) {$\psi$};
        \node [style=none] (22) at (-7.25, -2.75) {};
    \end{pgfonlayer}
    \begin{pgfonlayer}{edgelayer}
        \draw (13.center) to (19.center);
        \draw (4) to (17.center);
        \draw (7) to (1.center);
        \draw[bend left=45, looseness=1.25] (5) to (13.center);
        \draw[bend right=45, looseness=1.25] (5) to (11.center);
        \draw (2.center) to (15);
        \draw (7) to (0.center);
        \draw (6) to (20.center);
        \draw (7) to (22.center);
        \draw (3.center) to (16);
    \end{pgfonlayer}
\end{tikzpicture}\]
 In the concrete basis structure of equation \ref{eq:concreteBS}), conjugation boils down to conjugating matrix entries when matrices are expressed in the corresponding orthonormal  basis.

In a $\dag$-SMC $\mathcal{C}$, any basis structure induces a bijection $\Lambda: \mathcal{C}(I,A) \rightarrow \mathcal{C}(A,A)$ between states and endomorphisms of $A$, depicted graphically as:
\[
\Lambda(\psi)=\ \ \ \raisebox{-9mm}{
\begin{tikzpicture}
    \begin{pgfonlayer}{nodelayer}
        \node [style=none] (0) at (4, -1) {};
        \node [style=none] (1) at (6.25, -1) {};
        \node [style=green dot] (2) at (4, -2) {$\psi$};
        \node [style=none] (3) at (5, -2) {$=$};
        \node [style=green dot] (4) at (6.25, -2) {};
        \node [style=green dot] (5) at (6.75, -2.5) {$\psi$};
        \node [style=none] (6) at (4, -3) {};
        \node [style=none] (7) at (5.5, -3) {};
    \end{pgfonlayer}
    \begin{pgfonlayer}{edgelayer}
        \draw (4) to (1.center);
        \draw (0.center) to (2);
        \draw (6.center) to (2);
        \draw (4) to (5);
        \draw (4) to (7.center);
    \end{pgfonlayer}
\end{tikzpicture}
}
\]
%\begin{equation}
%\def\JPicScale{0.6}
%\input{pictures/Lambda.pst}
%\end{equation}
In \textbf{FdHilb} $\Lambda(\psi)$ is unitary iff $\ket{\psi}$ is unbiased with respect to the basis $\{\ket{i}\}$ copied by $\delta$. Inspired by this, in the general setting a state $\psi$ is defined to be \emph{unbiased with respect to a basis structure $(A,\delta, \epsilon)$} iff $\Lambda(\psi)$ is \em unitary \em i.e.~the dagger and the inverse coincide.

States which are unbiased w.r.t. a particular basis structure are filled with the same colour as the basis structure. In the explicit case of qubits (i.e. the object $\mathbb{C}^2$ in \textbf{FdHilb}) the unbiased states are those of the form $\ket{e_1}+e^{i\alpha}\ket{e_2}$ where $\{\ket{e_i}\}$ is the copied basis, and we label these states with the phase, $\alpha$. For example, if the green basis structure corresponds to $\{\ket{0},\ket{1}\}$ we respectively depict $\ket{0}+e^{i\alpha}\ket{1}$ and $\Lambda(\ket{0}+e^{i\alpha}\ket{1})$ by:
\[
\begin{tikzpicture}
    \begin{pgfonlayer}{nodelayer}
        \node [style=none] (0) at (10, -1.75) {};
        \node [style=none] (1) at (13, -1.75) {};
        \node [style=green dot] (2) at (10, -2.25) {$\alpha$};
        \node [style=green dot] (3) at (13, -2.25) {$\alpha$};
        \node [style=none] (4) at (13, -2.75) {};
    \end{pgfonlayer}
    \begin{pgfonlayer}{edgelayer}
        \draw (2) to (0.center);
        \draw (3) to (4.center);
        \draw (3) to (1.center);
    \end{pgfonlayer}
\end{tikzpicture}
\]
%\begin{equation}
%\def\JPicScale{0.6}
%\input{pictures/alpha.pst}
%\end{equation}

It's easy to show that the unitary $\Lambda(\psi)$s form a group. In the case of any basis structure on the object $\mathbb{C}^2$ in \textbf{FdHilb} (i.e. the qubit case), these unitaries are exactly the phase rotations w.r.t.~the copied basis; for this reason the group formed by the $\Lambda(\psi)$ unitaries is termed the \emph{phase group} \cite{RedGreen}.

 In quantum mechanics the relationship between \emph{different} orthonormal bases is clearly of crucial importance -- they represent incompatible observables. Work has been done on abstractly characterising \emph{mutually unbiased} basis structures \cite{RedGreen} -- those corresponding to bases which are unbiased w.r.t.~one another. It was shown that the basis elements of one basis structure are unbiased w.r.t.~the other basis structure -i.e.~imitating  the concrete definition of unbiased bases- if and only if \cite{RedGreen}:
 \[
\begin{tikzpicture}
    \begin{pgfonlayer}{nodelayer}
        \node [style=none] (0) at (-3.5, -1.25) {};
        \node [style=none] (1) at (-1.75, -1.25) {};
        \node [style=green dot] (2) at (-3.5, -1.75) {};
        \node [style=green dot] (3) at (-1.75, -1.75) {};
        \node [style=none] (4) at (-2.5, -2.25) {$\stackrel{H}{=}$};
        \node [style=red dot] (5) at (-3.5, -2.75) {};
        \node [style=red dot] (6) at (-1.75, -2.75) {};
        \node [style=none] (7) at (-3.5, -3.25) {};
        \node [style=none] (8) at (-1.75, -3.25) {};
    \end{pgfonlayer}
    \begin{pgfonlayer}{edgelayer}
        \draw (5) to (7.center);
        \draw (1.center) to (3);
        \draw[bend right=45, looseness=1.50] (2) to (5);
        \draw (0.center) to (2);
        \draw[bend right=315, looseness=1.50] (2) to (5);
        \draw (6) to (8.center);
    \end{pgfonlayer}
\end{tikzpicture}
 \]
where conventionally we denote one such basis structure with green dots and the other with red dots.
As also shown in \cite{RedGreen}, certain importaint pairs of mutually unbiased bases in quantum theory (e.g.~the Z and X spin observables) obey  a strictly stronger set of equations,\footnote{That is, strictly stronger provided that we are considering basis structures rather than a monoid-comonoid pair.}  namely those that define up-to-scalar-multiples a so-called bialgebra \cite{StreetBook}:
%\begin{equation}
%\def\JPicScale{0.6}
%\input{pictures/bialgebra.pst}
%\end{equation}
\[
\begin{tikzpicture}[scale=0.8]
    \begin{pgfonlayer}{nodelayer}
        \node [style=none] (0) at (-6, -0.75) {};
        \node [style=none] (1) at (-5, -0.75) {};
        \node [style=none] (2) at (-3, -0.75) {};
        \node [style=none] (3) at (-2, -0.75) {};
        \node [style=green dot] (4) at (-6, -1.25) {};
        \node [style=green dot] (5) at (-5, -1.25) {};
        \node [style=red dot] (6) at (-2.5, -1.25) {};
        \node [style=none] (7) at (0.5, -1.25) {};
        \node [style=none] (8) at (1.5, -1.25) {};
        \node [style=none] (9) at (3.75, -1.25) {};
        \node [style=none] (10) at (4.75, -1.25) {};
        \node [style=none] (11) at (7.25, -1.25) {};
        \node [style=none] (12) at (8.25, -1.25) {};
        \node [style=none] (13) at (10.5, -1.25) {};
        \node [style=none] (14) at (11.5, -1.25) {};
        \node [style=none] (15) at (-3.75, -1.75) {$\stackrel{B_1}{=}$};
        \node [style=red dot] (16) at (1, -1.75) {};
        \node [style=none] (17) at (2.5, -1.75) {$\stackrel{B_2}{=}$};
        \node [style=green dot] (18) at (7.75, -1.75) {};
        \node [style=none] (19) at (9.25, -1.75) {$\stackrel{B_3}{=}$};
        \node [style=green dot] (20) at (3.75, -2) {};
        \node [style=green dot] (21) at (4.75, -2) {};
        \node [style=red dot] (22) at (10.5, -2) {};
        \node [style=red dot] (23) at (11.5, -2) {};
        \node [style=red dot] (24) at (-6, -2.25) {};
        \node [style=red dot] (25) at (-5, -2.25) {};
        \node [style=green dot] (26) at (-2.5, -2.5) {};
        \node [style=green dot] (27) at (1, -2.5) {};
        \node [style=red dot] (28) at (7.75, -2.5) {};
        \node [style=none] (29) at (-6, -2.75) {};
        \node [style=none] (30) at (-5, -2.75) {};
        \node [style=none] (31) at (-3, -3) {};
        \node [style=none] (32) at (-2, -3) {};
    \end{pgfonlayer}
    \begin{pgfonlayer}{edgelayer}
        \draw (11.center) to (18);
        \draw (3.center) to (6);
        \draw (25) to (4);
        \draw (14.center) to (23);
        \draw (9.center) to (20);
        \draw (10.center) to (21);
        \draw (13.center) to (22);
        \draw (32.center) to (26);
        \draw (12.center) to (18);
        \draw (25) to (30.center);
        \draw (0.center) to (4);
        \draw (26) to (31.center);
        \draw (1.center) to (5);
        \draw (18) to (28);
        \draw (6) to (26);
        \draw (5) to (24);
        \draw (24) to (29.center);
        \draw (2.center) to (6);
        \draw (16) to (27);
        \draw (5) to (25);
        \draw (4) to (24);
        \draw (7.center) to (16);
        \draw (8.center) to (16);
    \end{pgfonlayer}
\end{tikzpicture}
\]
below we will omit specifying `up-to-scalar-multiples' when referring to these laws.

The key feature of the graphical calculus is that particular behaviors correspond to radical changes of the topology of the picture e.g.~being `an eigenstate' or being `mutually unbiased' both correspond to pictures decomposing into disconnected components - cf.~equations ($C_1$) and ($H$) respectively.  This is the very heart of categorical quantum mechanics: essential concepts are expressed in a language for which all can be reduced to tensor product structure, `disconnecting' then standing for `disentanglement'.  It goes without saying that topological distinctions come with clear behavioral differences.   In Section \ref{sec:topoclass} we will classify tripartite entanglement also according to this paradigm.

%\bR Now consider the following morphism \hfill (WAS IN BILL"S SOURCE)\e

\section{GHZ and W states represented graphically}\label{GHZWgraphicallySec}

We now move to consider the concrete case of qubits. We will be using two mutually unbiased basis structures, $\Delta_Z = (\mathbb{C}^2, \delta_Z, \epsilon_Z)$ which corresponds to the $\ket{0},\ket{1}$ basis and which will be represented graphically by green dots, and $\Delta_X = (\mathbb{C}^2, \delta_X, \epsilon_X)$ which corresponds to the $\ket{+},\ket{-}$ basis and which will be represented graphically by red dots. Viewed as a tripartite state $\mathbb{C}\rightarrow \mathbb{C}^2\otimes\mathbb{C}^2\otimes\mathbb{C}^2$ the green $\delta_Z$ is in fact the GHZ state. Thus the GHZ state can be depicted graphically as:

\begin{equation}\label{GHZ}
\begin{tikzpicture}[scale=0.7]
    \begin{pgfonlayer}{nodelayer}
        \node [style=none] (0) at (4.5, 1) {};
        \node [style=green dot] (1) at (4.5, 0) {};
        \node [style=none] (2) at (3.75, -0.75) {};
        \node [style=none] (3) at (5.25, -0.75) {};
    \end{pgfonlayer}
    \begin{pgfonlayer}{edgelayer}
        \draw (0.center) to (1);
        \draw (1) to (2.center);
        \draw (3.center) to (1);
    \end{pgfonlayer}
\end{tikzpicture}
\end{equation}

The same diagram but with a red dot is also a GHZ class state: it's the state $\ket{+++}+\ket{---}$ which is clearly obtained from the standard GHZ state via local basis transformations.
Since any state in the GHZ class is related to the GHZ state via local linear operations (recall equation \ref{sloccconditionEq}), such states can be depicted as:

\begin{equation}\label{GHZclass}
\begin{tikzpicture}[scale=0.8]
    \begin{pgfonlayer}{nodelayer}
        \node [style=none] (0) at (4.5, 1.75) {};
        \node [style=white dot] (1) at (4.5, 1) {$A_3$};
        \node [style=green dot] (2) at (4.5, 0) {};
        \node [style=white dot] (3) at (3.75, -0.75) {$A_1$};
        \node [style=white dot] (4) at (5.25, -0.75) {$A_2$};
        \node [style=none] (5) at (3.25, -1.25) {};
        \node [style=none] (6) at (5.75, -1.25) {};
    \end{pgfonlayer}
    \begin{pgfonlayer}{edgelayer}
        \draw (2) to (5.center);
        \draw (0.center) to (2);
        \draw (6.center) to (2);
    \end{pgfonlayer}
\end{tikzpicture}
\end{equation}
where $A_1$, $A_2$ and $A_3$ are linear maps. Now consider the following diagram:
\begin{equation}\label{W}
\begin{tikzpicture}[scale=0.65]
    \begin{pgfonlayer}{nodelayer}
        \node [style=none] (0) at (4.5, 2.25) {};
        \node [style=red dot] (1) at (4.5, 1.5) {};
        \node [style=green dot] (2) at (3.75, 0.25) {${\pi\over 3}$};
        \node [style=green dot] (3) at (5.25, 0.25) {${\pi\over 3}$};
        \node [style=red dot] (4) at (3, -1) {};
        \node [style=green dot] (5) at (4.5, -1) {${\pi\over 3}$};
        \node [style=red dot] (6) at (6, -1) {};
        \node [style=none] (7) at (2.25, -1.5) {};
        \node [style=none] (8) at (6.75, -1.5) {};
    \end{pgfonlayer}
    \begin{pgfonlayer}{edgelayer}
        \draw (6) to (8.center);
        \draw (7.center) to (4);
        \draw (1) to (4);
        \draw (4) to (6);
        \draw (1) to (6);
        \draw (0.center) to (1);
    \end{pgfonlayer}
\end{tikzpicture}
\end{equation}
When evaluated in \textbf{FdHilb} (i.e. concretely composing the linear maps represented by the graphical components), and ignoring global phase and normalisation, we get the W state.

Comparing the two diagrams \ref{GHZ} and \ref{W} it is striking to see how they seem to embody the global/local entanglement distinction discussed in section \ref{3qubitbackgroundSec}. In the GHZ state, \ref{GHZ}, the three systems are all connected together via the green dot, mirroring the genuine global tripartite entanglement which they share in this state. In contrast, in the W state, \ref{W}, the systems are connected in a pairwise fashion, mirroring the pairwise entanglement in this state.

If we change the phases, from $\pi / 3$ to $0$ then we would wind up with a GHZ-class state:
\begin{equation}\label{spider}
\begin{tikzpicture}[scale=0.5]
    \begin{pgfonlayer}{nodelayer}
        \node [style=none] (0) at (4.5, 2.25) {};
        \node [style=red dot] (1) at (4.5, 1.5) {};
        \node [style=none] (2) at (9.25, 1.5) {};
        \node [style=none] (3) at (7.25, 0.5) {$=$};
        \node [style=red dot] (4) at (9.25, 0.5) {};
        \node [style=none] (5) at (8.5, -0.25) {};
        \node [style=none] (6) at (10, -0.25) {};
        \node [style=red dot] (7) at (3, -1) {};
        \node [style=red dot] (8) at (6, -1) {};
        \node [style=none] (9) at (2.25, -1.5) {};
        \node [style=none] (10) at (6.75, -1.5) {};
    \end{pgfonlayer}
    \begin{pgfonlayer}{edgelayer}
        \draw (4) to (6.center);
        \draw (2.center) to (4);
        \draw (8) to (10.center);
        \draw (9.center) to (7);
        \draw (0.center) to (1);
        \draw (7) to (8);
        \draw (4) to (5.center);
        \draw (1) to (8);
        \draw (1) to (7);
    \end{pgfonlayer}
\end{tikzpicture}\end{equation}
So depending on the choice of phases we may end up with a W-class or a GHZ-class state. For
the general case:
\begin{equation}\label{alphabetagamma}
\begin{tikzpicture}[scale=0.6]
    \begin{pgfonlayer}{nodelayer}
        \node [style=none] (0) at (4.5, 2.25) {};
        \node [style=red dot] (1) at (4.5, 1.5) {};
        \node [style=green dot] (2) at (3.75, 0.25) {$\alpha$};
        \node [style=green dot] (3) at (5.25, 0.25) {$\beta$};
        \node [style=red dot] (4) at (3, -1) {};
        \node [style=green dot] (5) at (4.5, -1) {$\gamma$};
        \node [style=red dot] (6) at (6, -1) {};
        \node [style=none] (7) at (2.25, -1.5) {};
        \node [style=none] (8) at (6.75, -1.5) {};
    \end{pgfonlayer}
    \begin{pgfonlayer}{edgelayer}
        \draw (0.center) to (1);
        \draw (6) to (8.center);
        \draw (1) to (4);
        \draw (7.center) to (4);
        \draw (1) to (6);
        \draw (4) to (6);
    \end{pgfonlayer}
\end{tikzpicture}
\end{equation}
we will now verify for which  values of $\alpha$, $\beta$ and $\gamma$ this state is a GHZ-class state, and for which it is a W-class state.  Concretely composing linear maps gives:
\begin{equation}
\ket{\psi}=(1 + e^{i(\alpha + \beta + \gamma)})\ket{000} + (e^{i\alpha} + e^{i(\beta + \gamma)})\ket{011} + (e^{i\beta} + e^{i(\alpha + \gamma)})\ket{101} + (e^{i\gamma} + e^{i(\alpha + \beta)})\ket{110}
\end{equation}
for which   the 3-tangle is equal to:
\begin{equation}
\tau_{ABC} = 16 |a| |b| |c| |d|
\end{equation}
where
\beq
a =1 + e^{i(\alpha + \beta + \gamma)}\quad
b =e^{i\gamma} + e^{i(\alpha + \beta)}\quad
c =e^{i\alpha} + e^{i(\beta + \gamma)}\quad
d =e^{i\beta} + e^{i(\alpha + \gamma)}
\eeq
Thus, the tangle is zero (and $\ket{\psi}$ not a GHZ-class state) when one of the following conditions holds:
\begin{equation}\label{constraints}
\begin{array}{l}
\alpha + \beta + \gamma = \pi\\
\gamma - \alpha - \beta = \pi\\
\alpha - \beta - \gamma = \pi\\
\beta - \alpha - \gamma = \pi
\end{array}\end{equation}
%\section{Conclusions}
%Now, drawing together equations \ref{GHZclass}, \ref{redGHZ} and \ref{spider} we see that,
So unless one of these conditions holds, we must be able to find linear maps $A_1$, $A_2$, $A_3$, such that:

\begin{equation}
\begin{tikzpicture}[scale=0.7]
    \begin{pgfonlayer}{nodelayer}
        \node [style=none] (0) at (-5.75, 2) {};
        \node [style=none] (1) at (5.75, 2) {};
        \node [style=none] (2) at (0, 1.5) {};
        \node [style=red dot] (3) at (-5.75, 1.25) {};
        \node [style=white dot] (4) at (5.75, 1.25) {$A_3$};
        \node [style=white dot] (5) at (0, 0.75) {$A_3$};
        \node [style=red dot] (6) at (5.75, 0.25) {};
        \node [style=green dot] (7) at (-6.5, 0) {$\alpha$};
        \node [style=green dot] (8) at (-5, 0) {$\beta$};
        \node [style=none] (9) at (-2.5, 0) {$=$};
        \node [style=none] (10) at (2.5, 0) {$=$};
        \node [style=red dot] (11) at (0, -0.25) {};
        \node [style=red dot] (12) at (5.25, -0.5) {};
        \node [style=red dot] (13) at (6.25, -0.5) {};
        \node [style=white dot] (14) at (-0.75, -1) {$A_1$};
        \node [style=white dot] (15) at (0.75, -1) {$A_2$};
        \node [style=red dot] (16) at (-7.25, -1.25) {};
        \node [style=green dot] (17) at (-5.75, -1.25) {$\gamma$};
        \node [style=red dot] (18) at (-4.25, -1.25) {};
        \node [style=white dot] (19) at (4.25, -1.25) {$A_1$};
        \node [style=white dot] (20) at (7.25, -1.25) {$A_2$};
        \node [style=none] (21) at (-1.25, -1.5) {};
        \node [style=none] (22) at (1.25, -1.5) {};
        \node [style=none] (23) at (-8, -1.75) {};
        \node [style=none] (24) at (-3.5, -1.75) {};
        \node [style=none] (25) at (3.5, -1.75) {};
        \node [style=none] (26) at (8, -1.75) {};
    \end{pgfonlayer}
    \begin{pgfonlayer}{edgelayer}
        \draw (16) to (18);
        \draw (3) to (16);
        \draw (18) to (24.center);
        \draw (25.center) to (12);
        \draw (0.center) to (3);
        \draw (11) to (21.center);
        \draw (22.center) to (11);
        \draw (6) to (12);
        \draw (12) to (13);
        \draw (23.center) to (16);
        \draw (6) to (13);
        \draw (3) to (18);
        \draw (2.center) to (11);
        \draw (13) to (26.center);
        \draw (1.center) to (6);
    \end{pgfonlayer}
\end{tikzpicture}
\end{equation}
i.e. we should be able to `pull out' the three phases, in the process transforming them into linear maps, leaving a free central triangle, which can then be closed down to a single red dot using the spider law. In the case where one of the four conditions does hold, something stops us from pulling the phases out, and they cause a `log jam', preventing us from closing down the triangle to a dot, and dooming the state to be without genuine global tripartite entanglement.

\section{GHZ and W states analyzed graphically}\label{sec:topoclass}

\newcommand{\rd}{\bR\bullet\e}
%\begin{tikzpicture}
%   \begin{pgfonlayer}{nodelayer}
%       \node [style=red dot] (0) at (0, 0) {};
%   \end{pgfonlayer}
%\end{tikzpicture}
%}

\newcommand{\rpt}{
\raisebox{-1.8mm}{
\begin{tikzpicture}[scale=0.44]
    \begin{pgfonlayer}{nodelayer}
        \node [style=none] (0) at (0, 0.5) {};
        \node [style=red dot] (1) at (0, -0.25) {};
    \end{pgfonlayer}
    \begin{pgfonlayer}{edgelayer}
        \draw (1) to (0.center);
    \end{pgfonlayer}
\end{tikzpicture}
}}

\newcommand{\rppt}{
\raisebox{-1.8mm}{
\begin{tikzpicture}[scale=0.44]
    \begin{pgfonlayer}{nodelayer}
        \node [style=none] (0) at (0, 0.5) {};
        \node [style=red dot] (1) at (0, -0.25) {\small$\!\pi\!$};
    \end{pgfonlayer}
    \begin{pgfonlayer}{edgelayer}
        \draw (1) to (0.center);
    \end{pgfonlayer}
\end{tikzpicture}
}}

\newcommand{\gbs}{\bigl(\!\!
\raisebox{-2.3mm}{
\begin{tikzpicture}[scale=0.44]
    \begin{pgfonlayer}{nodelayer}
        \node [style=none] (0) at (2.75, -1.5) {};
        \node [style=none] (1) at (3.75, -1.5) {};
        \node [style=green dot] (2) at (3.25, -2) {};
        \node [style=green dot] (3) at (4.75, -2) {};
        \node [style=none] (4) at (4, -2.25) {$,$};
        \node [style=none] (5) at (3.25, -2.75) {};
        \node [style=none] (6) at (4.75, -2.75) {};
    \end{pgfonlayer}
    \begin{pgfonlayer}{edgelayer}
        \draw (2) to (0.center);
        \draw (2) to (5.center);
        \draw (2) to (1.center);
        \draw (3) to (6.center);
    \end{pgfonlayer}
\end{tikzpicture}
}\!\bigr)}

\newcommand{\gbr}{\bigl(\!\!
\raisebox{-2.3mm}{
\begin{tikzpicture}[scale=0.44]
    \begin{pgfonlayer}{nodelayer}
        \node [style=none] (0) at (2.75, -1.5) {};
        \node [style=none] (1) at (3.75, -1.5) {};
        \node [style=red dot] (2) at (3.25, -2) {};
        \node [style=red dot] (3) at (4.75, -2) {};
        \node [style=none] (4) at (4, -2.25) {$,$};
        \node [style=none] (5) at (3.25, -2.75) {};
        \node [style=none] (6) at (4.75, -2.75) {};
    \end{pgfonlayer}
    \begin{pgfonlayer}{edgelayer}
        \draw (2) to (0.center);
        \draw (2) to (5.center);
        \draw (2) to (1.center);
        \draw (3) to (6.center);
    \end{pgfonlayer}
\end{tikzpicture}
}\!\bigr)}

In this section we wish to produce the constraints (\ref{constraints}) not by direct computation but within the diagrammatic language.  We will show how these four constraints can be classified.
% in terms of classical points, and that  the +'s and -'s will have a clear structural meaning.

First observe that:
\[
\begin{tikzpicture}
    \begin{pgfonlayer}{nodelayer}
        \node [style=none] (0) at (-3.5, -0.75) {};
        \node [style=none] (1) at (-0.25, -0.75) {};
        \node [style=none] (2) at (3.75, -0.75) {};
        \node [style=none] (3) at (8, -0.75) {};
        \node [style=none] (4) at (10.5, -0.75) {};
        \node [style=red dot] (5) at (-3.5, -1.25) {};
        \node [style=red dot] (6) at (-0.25, -1.25) {};
        \node [style=green dot] (7) at (3.75, -1.5) {};
        \node [style=green dot] (8) at (8, -1.5) {};
        \node [style=red dot] (9) at (7, -1.75) {};
        \node [style=green dot] (10) at (10.5, -1.75) {};
        \node [style=green dot] (11) at (-4, -2) {$\xi$};
        \node [style=green dot] (12) at (-3, -2) {$\zeta$};
        \node [style=none] (13) at (-2.25, -2) {$=$};
        \node [style=green dot] (14) at (-1.5, -2) {$\xi$};
        \node [style=green dot] (15) at (-0.75, -2) {};
        \node [style=green dot] (16) at (0.25, -2) {};
        \node [style=green dot] (17) at (1, -2) {$\zeta$};
        \node [style=none] (18) at (1.75, -2) {$\stackrel{B_1}{=}$};
        \node [style=green dot] (19) at (2.5, -2) {$\xi$};
        \node [style=red dot] (20) at (3.5, -2) {};
        \node [style=green dot] (21) at (5, -2) {$\zeta$};
        \node [style=none] (22) at (5.75, -2) {$=$};
        \node [style=none] (23) at (8.75, -2) {$=$};
        \node [style=green dot] (24) at (6.5, -2.5) {$\xi$};
        \node [style=green dot] (25) at (7.5, -2.5) {$\zeta$};
        \node [style=white dot] (26) at (9.75, -2.5) {\!\!$\xi\rd\zeta$\!\!};
        \node [style=red dot] (27) at (-3.5, -2.75) {};
        \node [style=red dot] (28) at (-0.25, -2.75) {};
        \node [style=none] (29) at (-3.5, -3.25) {};
        \node [style=none] (30) at (-0.25, -3.25) {};
        \node [style=none] (31) at (3.75, -3.25) {};
        \node [style=none] (32) at (8, -3.25) {};
        \node [style=none] (33) at (10.5, -3.25) {};
    \end{pgfonlayer}
    \begin{pgfonlayer}{edgelayer}
        \draw[bend right=45, looseness=1.50] (5) to (27);
        \draw (8) to (9);
        \draw (4.center) to (10);
        \draw (27) to (29.center);
        \draw (10) to (33.center);
        \draw (19) to (20);
        \draw (14) to (15);
        \draw (8) to (32.center);
        \draw (16) to (17);
        \draw[bend right=315, looseness=1.50] (6) to (28);
        \draw (3.center) to (8);
        \draw (2.center) to (7);
        \draw (20) to (21);
        \draw (1.center) to (6);
        \draw (7) to (31.center);
        \draw (7) to (20);
        \draw[bend right=45, looseness=1.50] (6) to (28);
        \draw (24) to (9);
        \draw (9) to (25);
        \draw (0.center) to (5);
        \draw[bend right=315, looseness=1.50] (5) to (27);
        \draw (28) to (30.center);
        \draw (10) to (26);
    \end{pgfonlayer}
\end{tikzpicture}\]
where we set $\xi\rd\zeta:=\delta_X\circ(\xi\otimes\zeta)$. It then follows by equation $C_1$ that:

\begin{proposition}\label{MainProp}
If basis structures $\gbr$ and $\gbs$ form a bialgebra then the endomorphism:
\[
\begin{tikzpicture}
    \begin{pgfonlayer}{nodelayer}
        \node [style=none] (0) at (-3.5, -0.75) {};
        \node [style=red dot] (1) at (-3.5, -1.25) {};
        \node [style=green dot] (2) at (-4, -2) {$\xi$};
        \node [style=green dot] (3) at (-3, -2) {$\zeta$};
        \node [style=red dot] (4) at (-3.5, -2.75) {};
        \node [style=none] (5) at (-3.5, -3.25) {};
    \end{pgfonlayer}
    \begin{pgfonlayer}{edgelayer}
        \draw (0.center) to (1);
        \draw[bend right=315, looseness=1.50] (1) to (4);
        \draw[bend right=45, looseness=1.50] (1) to (4);
        \draw (4) to (5.center);
    \end{pgfonlayer}
\end{tikzpicture}
\]
is disconnected whenever $\xi\rd\zeta$ is a basis element of $\gbs$.  Explicitly, up-to-a-scalar we obtain:
\[
\begin{tikzpicture}
    \begin{pgfonlayer}{nodelayer}
        \node [style=none] (0) at (-3.5, -0.75) {};
        \node [style=red dot] (1) at (-3.5, -1.5) {\!\!$\xi\rd\zeta$\!\!};
        \node [style=red dot] (2) at (-3.5, -2.75) {\!\!$\xi\rd\zeta$\!\!};
        \node [style=none] (3) at (-3.5, -3.5) {};
    \end{pgfonlayer}
    \begin{pgfonlayer}{edgelayer}
        \draw (0.center) to (1);
        \draw (2) to (3.center);
    \end{pgfonlayer}
\end{tikzpicture}\]
\end{proposition}

\begin{definition}\em
For basis structures $\gbr$ and $\gbs$ on $A$ which form a bialgebra we  call a pair of $\gbs$-phases $\xi, \zeta:I\to A$ \em supplementary \em when $i:=\xi\rd\zeta:I\to A$ is (up-to-a-scalar) a basis element of  $\gbr$. More specifically, we say that $\xi$ and $\zeta$ are \em $i$-supplementary\em.
\end{definition}

In the case of the concrete basis structures $\Delta_Z$ and $\Delta_X$, since  $\Delta_Z$ has two basis elements $\ket{0}$ and $\ket{1}$, there will be two kinds of supplementary.  We have
\beqa
\xi\rd\zeta
&=& \delta_Z\left((\ket{0}+ e^{i\xi}  \ket{1})\otimes (\ket{0}+ e^{i\zeta}  \ket{1}) \right)\\
&=& \delta_Z\left((\ket{00}+ e^{i(\xi+ \zeta)}  \ket{11})+( e^{i\xi}  \ket{01} + e^{i\zeta}  \ket{10})  \right)\\
&=& (1+ e^{i(\xi+ \zeta)} )\ket{0}+( e^{i\xi}   + e^{i\zeta} )\ket{1} \\
\eeqa
Hence $\xi\rd\zeta$ is either proportional to $\ket{0}$ and $\ket{1}$ respectively when:
\[
\xi+\zeta= \pi\mbox{\ \ $\Rightarrow$~$\ket{1}$-supplementary}\qquad\mbox{or}\qquad\xi+ \pi= \zeta \mbox{\ \ $\Rightarrow$~$\ket{0}$-supplementary}
\]
We can now reproduce equations (\ref{constraints}) in terms of `disconnectedness of a picture'.  We have:
\[
\begin{tikzpicture}[scale=1]
    \begin{pgfonlayer}{nodelayer}
        \node [style=none] (0) at (-2.5, 1.25) {};
        \node [style=none] (1) at (1.25, 1.25) {};
        \node [style=none] (2) at (5.25, 1.25) {};
        \node [style=red dot] (3) at (-2.5, 0.75) {};
        \node [style=red dot] (4) at (1.25, 0.75) {};
        \node [style=green dot] (5) at (5.25, 0.25) {};
        \node [style=green dot] (6) at (-3, 0) {$\beta$};
        \node [style=green dot] (7) at (-2, 0) {$\gamma$};
        \node [style=none] (8) at (-0.5, 0) {$=$};
        \node [style=green dot] (9) at (0.75, 0) {\scriptsize$\!\!\alpha+\beta\!\!$};
        \node [style=green dot] (10) at (1.75, 0) {$\gamma$};
        \node [style=none] (11) at (2.75, 0) {$=$};
        \node [style=green dot] (12) at (4.25, -0.5) {\scriptsize$\!(\alpha+\beta)\rd\gamma$\!};
        \node [style=red dot] (13) at (-3.5, -0.75) {};
        \node [style=green dot] (14) at (-2.5, -0.75) {$\alpha$};
        \node [style=red dot] (15) at (-1.5, -0.75) {};
        \node [style=red dot] (16) at (1.25, -0.75) {};
        \node [style=none] (17) at (-1, -1) {};
        \node [style=red dot] (18) at (-4.5, -1.25) {};
        \node [style=none] (19) at (1.25, -1.25) {};
        \node [style=none] (20) at (5.25, -1.25) {};
    \end{pgfonlayer}
    \begin{pgfonlayer}{edgelayer}
        \draw (3) to (13);
        \draw (0.center) to (3);
        \draw (16) to (19.center);
        \draw (3) to (15);
        \draw (17.center) to (15);
        \draw (2.center) to (5);
        \draw (13) to (18);
        \draw (5) to (20.center);
        \draw (15) to (13);
        \draw[bend right=315, looseness=1.50] (4) to (16);
        \draw (5) to (12);
        \draw[bend right=45, looseness=1.50] (4) to (16);
        \draw (1.center) to (4);
    \end{pgfonlayer}
\end{tikzpicture}
\]
so we obtain the two first of equations (\ref{constraints}); the remaining two equations  arise when  `plugging' $\rpt$ in the other corners of the W-state.  %Note here a clear difference between the first equation and the three others: for the first we obtain

Now, since by plugging $\rpt$ we obtain equations that are characteristic for the W class states, one expects that one can find a derivation of $\rpt$ from such a state.  This is indeed the case.    As shown in \cite{CK} also the W state can be endowed with a commutative Frobenius algebra structure, although this structure is not dagger, and more importantly, not special.   We still have some kind of a spider theorem, but now it also accounts for the number of loops in a picture, which has to be preserved.  It is indeed  `the value of the loop' that  is the characteristic difference between GHZ class states and W class states; we have:
\[
\raisebox{-7mm}{\begin{tikzpicture}
    \begin{pgfonlayer}{nodelayer}
        \node [style=none] (0) at (-4.75, -1.5) {};
        \node [style=none] (1) at (-3, -1.5) {};
        \node [style=dot] (2) at (-3, -1.75) {};
        \node [style=dot] (3) at (-4.75, -2.25) {};
        \node [style=none] (4) at (-4, -2.25) {=};
        \node [style=none] (5) at (-2, -2.25) {=};
        \node [style=dot] (6) at (-5.25, -2.5) {};
        \node [style=dot] (7) at (-3, -2.75) {};
        \node [style=none] (8) at (-4.75, -3) {};
        \node [style=none] (9) at (-3, -3) {};
    \end{pgfonlayer}
    \begin{pgfonlayer}{edgelayer}
        \draw[out=-45, in=-135, loop] (6) to ();
        \draw[bend left=315, looseness=1.50] (2) to (7);
        \draw (8.center) to (0.center);
        \draw[bend left=45, looseness=1.50] (2) to (7);
        \draw (6) to (3);
        \draw (7) to (9.center);
        \draw (1.center) to (2);
    \end{pgfonlayer}
\end{tikzpicture}}
\ \ \ \ \left\{
\begin{array}{cc}
\begin{tikzpicture}
    \begin{pgfonlayer}{nodelayer}
        \node [style=none] (0) at (-4.75, -1.5) {};
        \node [style=none] (1) at (-4.75, -3) {};
    \end{pgfonlayer}
    \begin{pgfonlayer}{edgelayer}
        \draw (0.center) to (1.center);
    \end{pgfonlayer}
\end{tikzpicture}
 & \raisebox{6mm}{\mbox{for the GHZ state}}\vspace{3mm}\\
\begin{tikzpicture}
    \begin{pgfonlayer}{nodelayer}
        \node [style=none] (0) at (0, 2) {};
        \node [style=dot] (1) at (0, 1.75) {};
        \node [style=dot] (2) at (0, 0.5) {};
        \node [style=none] (3) at (0, 0.25) {};
    \end{pgfonlayer}
    \begin{pgfonlayer}{edgelayer}
        \draw[out=45, in=135, loop] (2) to ();
        \draw (2) to (3.center);
        \draw (1) to (0.center);
        \draw[out=-45, in=-135, loop] (1) to ();
    \end{pgfonlayer}
\end{tikzpicture}
 & \raisebox{7mm}{\mbox{for the W state}}
\end{array}
\right.
\]
The commutative Frobenius algebra structure on W is given by:
\[
\left(\raisebox{-1.1cm}{\begin{tikzpicture}[scale=0.92]
    \begin{pgfonlayer}{nodelayer}
        \node [style=none] (0) at (0, 1) {};
        \node [style=none] (1) at (2, 1) {};
        \node [style=none] (2) at (4.25, 1) {};
        \node [style=red dot] (3) at (0, 0.5) {$\pi$};
        \node [style=green dot] (4) at (1, 0.5) {${\pi\over 3}$};
        \node [style=red dot] (5) at (2, 0.5) {$\pi$};
        \node [style=red dot] (6) at (4.25, 0.5) {$\pi$};
        \node [style=green dot] (7) at (0.5, -0.25) {${\pi\over 3}$};
        \node [style=green dot] (8) at (1.5, -0.25) {${\pi\over 3}$};
        \node [style=green dot] (9) at (3.75, -0.25) {${\pi\over 3}$};
        \node [style=green dot] (10) at (4.75, -0.25) {${\pi\over 3}$};
        \node [style=red dot] (11) at (7, -0.25) {};
        \node [style=none] (12) at (8.5, -0.25) {};
        \node [style=none] (13) at (2.5, -0.75) {$,$};
        \node [style=none] (14) at (6.25, -0.75) {$,$};
        \node [style=none] (15) at (7, -0.75) {};
        \node [style=none] (16) at (7.75, -0.75) {$,$};
        \node [style=red dot] (17) at (8.5, -0.75) {$\pi$};
        \node [style=red dot] (18) at (1, -1) {};
        \node [style=red dot] (19) at (3.25, -1) {};
        \node [style=green dot] (20) at (4.25, -1) {${\pi\over 3}$};
        \node [style=red dot] (21) at (5.25, -1) {};
        \node [style=none] (22) at (1, -1.5) {};
        \node [style=none] (23) at (3.25, -1.5) {};
        \node [style=none] (24) at (5.25, -1.5) {};
    \end{pgfonlayer}
    \begin{pgfonlayer}{edgelayer}
        \draw (22.center) to (18);
        \draw (18) to (5);
        \draw (21) to (24.center);
        \draw (0.center) to (3);
        \draw (2.center) to (6);
        \draw (5) to (1.center);
        \draw (6) to (21);
        \draw (3) to (5);
        \draw (23.center) to (19);
        \draw (6) to (19);
        \draw (18) to (3);
        \draw (12.center) to (17);
        \draw (15.center) to (11);
        \draw (19) to (21);
    \end{pgfonlayer}
\end{tikzpicture}}\right)
\]
and one can compute that:
\[
\begin{tikzpicture}[scale=0.92]
    \begin{pgfonlayer}{nodelayer}
        \node [style=none] (0) at (-0.25, 2.5) {};
        \node [style=red dot] (1) at (-0.25, 2) {$\pi$};
        \node [style=green dot] (2) at (-0.75, 1.25) {${\pi\over 3}$};
        \node [style=green dot] (3) at (0.25, 1.25) {${\pi\over 3}$};
        \node [style=none] (4) at (-3.5, 0.75) {};
        \node [style=dot] (5) at (-3.5, 0.5) {};
        \node [style=red dot] (6) at (-1.25, 0.5) {};
        \node [style=green dot] (7) at (-0.25, 0.5) {${\pi\over 3}$};
        \node [style=red dot] (8) at (0.75, 0.5) {};
        \node [style=none] (9) at (-5.5, 0.25) {};
        \node [style=none] (10) at (2.75, 0.25) {};
        \node [style=dot] (11) at (-5.5, 0) {};
        \node [style=none] (12) at (-4.5, 0) {$=$};
        \node [style=none] (13) at (-2.25, 0) {$=$};
        \node [style=none] (14) at (1.75, 0) {$=$};
        \node [style=red dot] (15) at (2.75, -0.25) {};
        \node [style=dot] (16) at (-3.5, -0.5) {};
        \node [style=red dot] (17) at (-1.25, -0.5) {$\pi$};
        \node [style=none] (18) at (-1.25, -0.5) {$\pi$};
        \node [style=green dot] (19) at (-0.25, -0.5) {${\pi\over 3}$};
        \node [style=red dot] (20) at (0.75, -0.5) {$\pi$};
        \node [style=none] (21) at (0.75, -0.5) {$\pi$};
        \node [style=dot] (22) at (-3.5, -1) {};
        \node [style=green dot] (23) at (-0.75, -1.25) {${\pi\over 3}$};
        \node [style=green dot] (24) at (0.25, -1.25) {${\pi\over 3}$};
        \node [style=red dot] (25) at (-0.25, -2) {};
        \node [style=red dot] (26) at (-0.25, -3) {$\pi$};
    \end{pgfonlayer}
    \begin{pgfonlayer}{edgelayer}
        \draw (26) to (25);
        \draw (18.center) to (6);
        \draw[bend left=45, looseness=1.50] (5) to (16);
        \draw (0.center) to (1);
        \draw (4.center) to (5);
        \draw (16) to (22);
        \draw[out=-45, in=-135, loop] (11) to ();
        \draw (11) to (9.center);
        \draw (25) to (20);
        \draw (1) to (6);
        \draw (8) to (21.center);
        \draw (17) to (20);
        \draw (25) to (17);
        \draw (1) to (8);
        \draw (15) to (10.center);
        \draw[bend left=315, looseness=1.50] (5) to (16);
        \draw (6) to (8);
    \end{pgfonlayer}
\end{tikzpicture}
\]
%Hence the fact that:
%\[
%\begin{tikzpicture}
%   \begin{pgfonlayer}{nodelayer}
%       \node [style=none] (0) at (4.25, 1.5) {};
%       \node [style=red dot] (1) at (4.25, 1) {$\pi$};
%       \node [style=red dot] (2) at (4.25, 0.5) {};
%       \node [style=green dot] (3) at (3.75, -0.25) {${\pi\over 3}$};
%       \node [style=green dot] (4) at (4.75, -0.25) {${\pi\over 3}$};
%       \node [style=red dot] (5) at (3.25, -1) {};
%       \node [style=green dot] (6) at (4.25, -1) {${\pi\over 3}$};
%       \node [style=red dot] (7) at (5.25, -1) {};
%       \node [style=none] (8) at (5.25, -1.5) {};
%       \node [style=red dot] (9) at (3.25, -1.75) {};
%   \end{pgfonlayer}
%   \begin{pgfonlayer}{edgelayer}
%       \draw (0.center) to (2);
%       \draw (9) to (5);
%       \draw (5) to (7);
%       \draw (2) to (5);
%       \draw (7) to (8.center);
%       \draw (2) to (7);
%   \end{pgfonlayer}
%\end{tikzpicture}
%\]
%is disconnected is indeed characteristic for the W state.
We can establish the last equality within graphical calculus by showing that, like RHS, also  LHS  is  orthogonal to $\rppt$.  Indeed:
\[
\begin{tikzpicture}[scale=0.92]
    \begin{pgfonlayer}{nodelayer}
        \node [style=red dot] (0) at (-6.5, 3) {$\pi$};
        \node [style=red dot] (1) at (-6.5, 2) {$\pi$};
        \node [style=red dot] (2) at (-2.5, 2) {};
        \node [style=green dot] (3) at (-7, 1.25) {${\pi\over 3}$};
        \node [style=green dot] (4) at (-6, 1.25) {${\pi\over 3}$};
        \node [style=green dot] (5) at (-3, 1.25) {${\pi\over 3}$};
        \node [style=green dot] (6) at (-2, 1.25) {${\pi\over 3}$};
        \node [style=red dot] (7) at (-7.5, 0.5) {};
        \node [style=green dot] (8) at (-6.5, 0.5) {${\pi\over 3}$};
        \node [style=red dot] (9) at (-5.5, 0.5) {};
        \node [style=red dot] (10) at (-3.5, 0.5) {};
        \node [style=green dot] (11) at (-2.5, 0.5) {${\pi\over 3}$};
        \node [style=red dot] (12) at (-1.5, 0.5) {};
        \node [style=red dot] (13) at (0.5, 0.5) {$\pi$};
        \node [style=red dot] (14) at (2.5, 0.5) {$\pi$};
        \node [style=none] (15) at (-4.5, 0) {$=$};
        \node [style=none] (16) at (-0.5, 0) {$\stackrel{Prop.~\ref{MainProp}}{=}$};
        \node [style=red dot] (17) at (-7.5, -0.5) {$\pi$};
        \node [style=none] (18) at (-7.5, -0.5) {$\pi$};
        \node [style=green dot] (19) at (-6.5, -0.5) {${\pi\over 3}$};
        \node [style=none] (20) at (-5.5, -0.5) {$\pi$};
        \node [style=red dot] (21) at (-5.5, -0.5) {$\pi$};
        \node [style=none] (22) at (-3.5, -0.5) {$\pi$};
        \node [style=red dot] (23) at (-3.5, -0.5) {$\pi$};
        \node [style=green dot] (24) at (-2.5, -0.5) {${\pi\over 3}$};
        \node [style=none] (25) at (-1.5, -0.5) {$\pi$};
        \node [style=red dot] (26) at (-1.5, -0.5) {$\pi$};
        \node [style=red dot] (27) at (0.5, -0.5) {$\pi$};
        \node [style=none] (28) at (0.5, -0.5) {$\pi$};
        \node [style=green dot] (29) at (1.5, -0.5) {${\pi\over 3}$};
        \node [style=none] (30) at (2.5, -0.5) {$\pi$};
        \node [style=red dot] (31) at (2.5, -0.5) {$\pi$};
        \node [style=none] (32) at (4.5, -0.5) {};
        \node [style=red dot] (33) at (4.5, -0.5) {};
        \node [style=green dot] (34) at (5.5, -0.5) {${\pi\over 3}$};
        \node [style=none] (35) at (6.5, -0.5) {};
        \node [style=red dot] (36) at (6.5, -0.5) {};
        \node [style=green dot] (37) at (8.5, -0.5) {$\pi$};
        \node [style=green dot] (38) at (-7, -1.25) {${\pi\over 3}$};
        \node [style=green dot] (39) at (-6, -1.25) {${\pi\over 3}$};
        \node [style=green dot] (40) at (-3, -1.25) {${\pi\over 3}$};
        \node [style=green dot] (41) at (-2, -1.25) {${\pi\over 3}$};
        \node [style=green dot] (42) at (1, -1.25) {${\pi\over 3}$};
        \node [style=green dot] (43) at (2, -1.25) {${\pi\over 3}$};
        \node [style=none] (44) at (3.5, -1.25) {$=$};
        \node [style=green dot] (45) at (5, -1.25) {${\pi\over 3}$};
        \node [style=green dot] (46) at (6, -1.25) {${\pi\over 3}$};
        \node [style=none] (47) at (7.5, -1.25) {$=$};
        \node [style=red dot] (48) at (8.5, -1.25) {};
        \node [style=red dot] (49) at (-6.5, -2) {};
        \node [style=red dot] (50) at (-2.5, -2) {};
        \node [style=red dot] (51) at (1.5, -2) {};
        \node [style=red dot] (52) at (5.5, -2) {};
        \node [style=red dot] (53) at (8.5, -2.25) {$\pi$};
        \node [style=red dot] (54) at (-6.5, -3) {$\pi$};
        \node [style=red dot] (55) at (-2.5, -3) {$\pi$};
        \node [style=red dot] (56) at (1.5, -3) {$\pi$};
        \node [style=red dot] (57) at (5.5, -3) {$\pi$};
    \end{pgfonlayer}
    \begin{pgfonlayer}{edgelayer}
        \draw (1) to (7);
        \draw (33) to (36);
        \draw (0) to (1);
        \draw (12) to (25.center);
        \draw (7) to (9);
        \draw (28.center) to (13);
        \draw (2) to (12);
        \draw (54) to (49);
        \draw[bend right=60] (37) to (48);
        \draw[bend left=45] (37) to (48);
        \draw (51) to (31);
        \draw (49) to (17);
        \draw (23) to (26);
        \draw (52) to (33);
        \draw (22.center) to (10);
        \draw (50) to (26);
        \draw (55) to (50);
        \draw (17) to (21);
        \draw (53) to (48);
        \draw (56) to (51);
        \draw (27) to (31);
        \draw (52) to (36);
        \draw (10) to (12);
        \draw (57) to (52);
        \draw (50) to (23);
        \draw (49) to (21);
        \draw (18.center) to (7);
        \draw (51) to (27);
        \draw (1) to (9);
        \draw (2) to (10);
        \draw (14) to (30.center);
        \draw (9) to (20.center);
    \end{pgfonlayer}
\end{tikzpicture}
\]
and a $\pi$-gate of one color within a loop-point of the other color is always $0$.

\section{Outlook}

We showed that W class states occur when a singularity takes place in the graphical calculus: plugging with a certain point `special' structural value leeds to disconnectedness.  The methods in this paper for analyzing the threepartite qubit states admit extrapolation to multiple qubits; for example, for:
\[
\begin{tikzpicture}[scale=0.6]
    \begin{pgfonlayer}{nodelayer}
        \node [style=none] (0) at (-3.5, 1.5) {};
        \node [style=none] (1) at (-0.5, 1.5) {};
        \node [style=red dot] (2) at (-3, 1) {};
        \node [style=green dot] (3) at (-2, 1) {$\delta$};
        \node [style=red dot] (4) at (-1, 1) {};
        \node [style=green dot] (5) at (-3, 0) {$\alpha$};
        \node [style=green dot] (6) at (-1, 0) {$\gamma$};
        \node [style=red dot] (7) at (-3, -1) {};
        \node [style=green dot] (8) at (-2, -1) {$\beta$};
        \node [style=red dot] (9) at (-1, -1) {};
        \node [style=none] (10) at (-3.5, -1.5) {};
        \node [style=none] (11) at (-0.5, -1.5) {};
    \end{pgfonlayer}
    \begin{pgfonlayer}{edgelayer}
        \draw (9) to (11.center);
        \draw (7) to (9);
        \draw (2) to (7);
        \draw (10.center) to (7);
        \draw (4) to (9);
        \draw (2) to (4);
        \draw (0.center) to (2);
        \draw (1.center) to (4);
    \end{pgfonlayer}
\end{tikzpicture}
\]
by plugging  two corners,  analogous singularities occur when either  $(\alpha + \beta) \rd (\gamma +\delta)$ or $\alpha  \rd (\beta + \gamma +\delta)$ are basis elements for $\gbs$. In this case the status of $\rpt$ isn't as clear as in the previous section where we could rely on the results of \cite{CK}, and hence requires further investigation.
On the other hand, while the results in \cite{CK} are restricted to the highly symmetrical so-called Frobenius states, the methods presented here don't have such a constraint, and may lead to an algebraic account on more general states than tripartite ones.

\bibliography{MPE_bob_bill}

\end{document}